\documentclass[12pt,a4paper]{article}
\usepackage{amsmath,amsfonts,latexsym,amssymb}
\usepackage[dvips]{graphics}

\newcommand{\bea}{\begin{eqnarray*}}
\newcommand{\eea}{\end{eqnarray*}}

\newcommand{\bean}{\begin{eqnarray}}
\newcommand{\eean}{\end{eqnarray}}

\newcommand{\eps}{\epsilon}

\begin{document}

\title{Outer Trapped Surfaces In Vaidya Spacetimes}

\author{Ishai Ben--Dov\thanks{Electronic mail: \tt ibd@uchicago.edu} 
\\\it Enrico Fermi Institute and Department of Physics 
\\ \it University of Chicago \\ \it 5640 S. Ellis Avenue    
\\ \it Chicago, Illinois 60637-1433}
\maketitle

\begin{abstract}

It is proven that in Vaidya spacetimes of bounded total mass, the outer
boundary, in spacetime, of the region containing outer trapped surfaces, is
the event horizon. Further, it is shown that the region containing trapped
surfaces in these spacetimes does not always extend to the event horizon.

\end{abstract}


\section{Introduction}   \label{introduction}

There has been renewed interest in the last several years in studying black
holes from a more local perspective, including attempting to come up with a
local definition for the boundaries of black holes.
A black hole is defined as a region in spacetime that cannot be observed from
infinity.
The boundary of the black hole region is the event horizon.
This is a 3-surface defined globally and therefore
knowledge of the entire future evolution of the spacetime is required
before the event horizon's position and even existence are known.
Turning next to local surfaces, the notion of an outer trapped surface was
introduced in the studies of black holes.
An outer trapped surface is a compact spacelike surface where the outgoing
null geodesics orthogonal to the surface are initially converging, i.e.\ their
expansion is negative.
Examples of outer trapped surfaces are 2-spheres inside the event horizon in
a Schwarzschild spacetime.
Verifying that a surface is outer trapped only requires knowledge of the
spacetime in a neighborhood of the surface.
In this sense outer trapped surfaces are local while the event horizon, as
mentioned, is global.

Assuming reasonable energy conditions and a non-nakedly singular spacetime,
outer trapped surfaces lie entirely inside the event horizon
\cite{HawkingEllis}.
In numerical applications locating these surfaces is of interest
in finding the existence of a black hole surrounding them, since locating the
event horizon is more difficult.
Outer trapped surfaces are also useful in excision techniques,
since outer trapped surfaces, and the region enclosed by
them, lie entirely inside the black hole. Hence, modifying this region
cannot affect the evolution outside the black hole and a numerical simulation
can be carried out more easily.

Consider some time slice of the spacetime, i.e.\ a spacelike 3-surface in the
spacetime.
One could consider the region in this 3-surface containing outer trapped
surfaces that lie entirely in this 3-surface.
The outer boundary of this region is the apparent horizon.
This is a spacelike 2-surface with the expansion of outgoing null geodesics
orthogonal to it vanishing \cite{HawkingEllis}. 
Given a foliation of the spacetime by spacelike
3-surfaces, the apparent horizon on each leaf of the foliation can be found,
and one can consider the union of the apparent horizons on all such leafs.
This is a 3-surface that will be referred to as the apparent 3-horizon, so as
to distinguish it from the apparent horizon, a 2-surface on one time slice.
In general, the apparent 3-horizon can be discontinuous.
Furthermore, the apparent horizons in spacetime are not unique.
A different foliation of the spacetime into spacelike surfaces can result in a
different location of the apparent horizon through the spacetime.
As a result the apparent 3-horizon is, in general,
neither unique, nor continuous.

More recently, the notions of trapping horizons \cite{Hayward94} and
dynamical horizons \cite{AshtekarKrishnan} were introduced.
These are 3-surfaces that, like the apparent horizon, are quasi-local
and therefore do not require the entire evolution of the spacetime in
order for them to be located.
Though these surfaces are smooth by definition, they come with additional
requirements.
It is not known, in general, whether these surfaces always exist in
spacetimes containing black holes.
Furthermore, these 3-surfaces need not be unique.\footnote{In
\cite{AshtekarGalloway}, Ashtekar and Galloway show that the intrinsic
structure of a dynamical horizon is unique, i.e.\ a 3-surface cannot admit
two distinct foliations both of which render the 3-surface a dynamical horizon.
However, they point out that there still remains freedom, so that, in general,
a spacetime may contain different dynamical horizons in the same region of
the spacetime.}

As described, the apparent horizon is defined by restricting attention to
outer trapped surfaces lying in a single time slice.
Consider removing this restriction.
Instead, consider the region \emph{in spacetime} containing outer
trapped surfaces. The outer boundary of this region is some unique
3-surface\footnote{It has not been proven that the outer boundary of this 
region is always sufficiently regular as to fully deserve the designation 
of 3-surface. The discussion here is heuristic as to illustrate and 
motivate the rigorous sections that follow.} that is certainly independent 
of any slicing of the spacetime. This 3-surface has not been studied much 
in the past.

Even in spherically symmetric spacetimes, where, due to symmetry, finding
apparent horizons on spherically symmetric slices is, relatively,
an easy task, finding this 3-surface is not as easy.
In spherically symmetric spacetimes, this 3-surface is spherically symmetric.
However, the outer trapped surfaces contained within the region
enclosed by this surface, need not be spherically symmetric.
In fact, they need not lie in spherically symmetric slices.
Not much is known about the locations of non-spherically symmetric outer
trapped surfaces, and this is the main source of difficulty in
locating this 3-surface.

Given some foliation of the spacetime, consider the apparent 3-horizon that
is associated with this foliation.
It is clear that the 3-surface, which is the boundary of the region containing
outer trapped surfaces, lies outside of (or coincides with) the
apparent 3-horizon and lies inside of (or coincides with) the event horizon.
However, in general, the apparent 3-horizon and the event horizon can be
separated in the dynamical regime, and therefore there is room for this
3-surface to lie somewhere in-between.

This point is well illustrated in Vaidya spacetimes, commonly used to describe
gravitational collapse that ends in the formation of a black hole.
The metric for a Vaidya spacetime is given by
\bean
\label{Metric}
g_{ab} = -\Big(1-\frac{R(v)}{r}\Big) dv_a dv_b + 2 dv_{(a} dr_{b)}+
r^2 \Big(d\theta_a d\theta_b + \sin^2\theta\, d\phi_a d\phi_b\Big)
\eean
with $R(v)$ some non-negative, non-decreasing, smooth function of $v$.
In this work, $R(v)$ is also assumed to be bounded from above with
a least upper bound, $R_0$.
The stress energy is given by
\bean
\label{Tmunu}
T_{ab} = \frac{R'(v)}{8\pi r^2} dv_a dv_b
\eean
The function $R(v)$ is non-negative to ensure a non-nakedly singular spacetime.
It is non-decreasing in order for the spacetime to satisfy the dominant energy
condition, as can be seen in (\ref{Tmunu}).
Finally, the additional requirement of it being bounded from above, is in
order to guarantee that the spacetime is asymptotically flat and asymptotes
to Schwarzschild.

This spherically symmetric spacetime describes the collapse of null dust in
forming a black hole of mass $M=\frac{1}{2} R_0$.
The metric is given in terms of the following coordinates.
Advanced time, $v$, areal radius, $r$, and the usual angular coordinates, 
$\theta$ and $\phi$.
In the special case $R(v)=2M$, with $M$ some positive constant, the metric is
that of Schwarzschild in ingoing Eddington-Finkelstein coordinates.

Given any 2-sphere, i.e.\ a surface of constant $v$ and $r$, the expansion,
$\Theta$, of outgoing null geodesics orthogonal to it is given by
\bean
\label{exp2sphere}
\Theta\,=\,-\frac{2\big(R(v)-r\big)}{r^2}
\eean
Since the apparent horizon on any spherically symmetric slice of this spacetime
is a 2-sphere, then it follows from (\ref{exp2sphere}) that the apparent
horizon on any spherically symmetric asymptotically flat slice
is the outermost 2-sphere satisfying $r=R(v)$ in that slice.
Consider the spacelike 3-surface given by $r=R(v)$ for all $v$.
Outside the $r=R(v)$ surface, 2-spheres are not outer trapped,
while inside it, 2-spheres are outer trapped.
The event horizon does not, in general, coincide with this 3-surface,
and thus there exists, in general, a non-empty region between the
surface $r=R(v)$ and the event horizon.

\begin{figure}
      \begin{center}
    \resizebox{13cm}{!}{\includegraphics{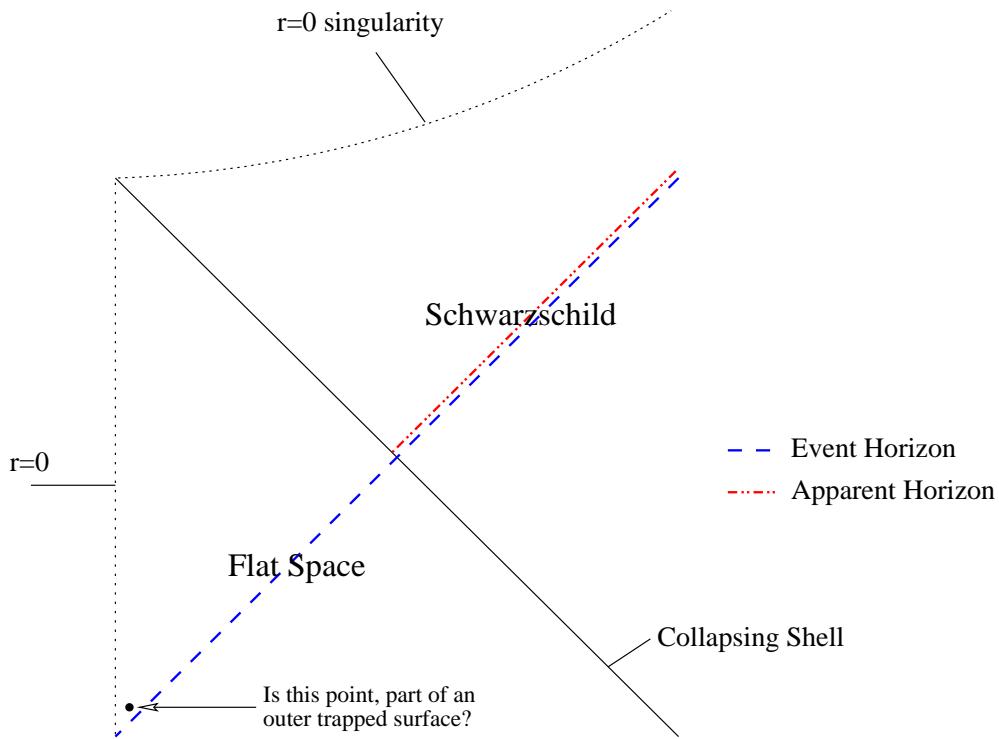}}
  \end{center}
  \caption{A spacetime diagram of the collapse of a thin null dust shell in
flat space. Two angular dimensions are suppressed. Points in this diagram are,
therefore, 2-spheres. The region to the left of the shell is a portion of flat
space. The region to the right of it is a portion of a Schwarzschild spacetime.
The event horizon, a null 3-surface, is shown. The point on
the left represents a single event (as part of a 2-sphere). It lies inside
the event horizon but is deep in the flat region. Is there an outer trapped
surface that contains this point?}
  \label{ShellFig}
\end{figure}

Moreover, notice that, in general, far in the past $R(v)$ may vanish and that
region of the spacetime will be a portion of flat space. In fact, the spacetime
where $R(v)=0$ for $v < v_0$ and $R(v)=2M$ for $v \geq v_0$, which can be
obtained as a limit of a smooth family of such Vaidya spacetimes, describes
the collapse of a thin null dust shell in flat space. A spacetime diagram
of a thin null dust shell collapse is shown in Fig.\ \ref{ShellFig}.

As can be seen in this figure, the event horizon extends into the flat
region. But, it is known that there are no outer trapped surfaces in flat
space.\footnote{This follows since, as mentioned, outer trapped surfaces
lie entirely inside an event horizon when certain conditions - which flat space
certainly satisfies - hold. Since flat space does not contain any black holes,
it therefore cannot contain any outer trapped surfaces either.}
Indeed, the surface $r=R(v)$ does not extend into the flat region
of the spacetime.
What about the boundary of the region containing outer trapped surfaces?
Could this 3-surface extend into the flat portion?

Eardley \cite{Eardley} conjectured that the boundary of the region that
contains marginally outer trapped surfaces coincides with the event horizon.
For this conjecture to be true, there need to exist marginally outer trapped
surfaces close to the event horizon in the flat portion.
For example, a marginally outer trapped surface must pass through the point
shown in Fig.\ \ref{ShellFig}.
However, as was just pointed, there are no marginally outer trapped surfaces
in flat space.
Could there be marginally outer trapped surfaces in this spacetime that lie,
in part, in the flat portion?

In \cite{Eardley}, Eardley gave an argument showing how one can get outer
trapped surfaces, parts of which extend beyond the apparent horizon on some
given slice.
He then discussed how one might, in spacetimes such as the null shell collapse,
find outer trapped surfaces that reach into the flat portion.
The idea is to have a 2-surface that lies mostly far in the
future, with a thin tendril that is almost null and lies near the event
horizon.
The tendril reaches into the flat portion and this 2-surface with a
tendril is outer trapped.

The existence of non-spherically symmetric marginally trapped surfaces in
Vaidya spacetimes was investigated numerically by
Schnetter and Krishnan \cite{SchnetterKrishnan}.
For specific choices of $R(v)$ they located marginally trapped surfaces, i.e.\ 
surfaces where both expansions are non-positive, that penetrate into
part of the flat region of the Vaidya spacetime.

These results do not address Eardley's conjecture.
Eardley's conjecture is about the boundary of the region containing
(marginally) outer trapped surfaces, i.e.\ surfaces with no restriction on
the expansion of ingoing null geodesics orthogonal to the 2-surface.
In contrast, the surfaces found by Schnetter and Krishnan are marginally
trapped, i.e.\ they satisfy the additional requirement that the ingoing
expansion is everywhere non-positive.
However, since trapped surfaces are, of course, also outer trapped, then
Schnetter and Krishnan's results show numerically that it is possible to find
marginally outer trapped surfaces extending into the flat region of a
Vaidya spacetime.
Moreover, one can also consider the region containing trapped surfaces and its
outer boundary.
This boundary is some unique 3-surface that lies inside of (or coincides with)
the event horizon.\footnote{In fact, it lies inside of (or coincides with) the
outer boundary of the region containing outer trapped surfaces.}
The numerical results of Schnetter and Krishnan show that this 3-surface
can extend into the flat region of a Vaidya spacetime.

The main purpose of this work is to prove that in Vaidya spacetimes
there are outer trapped surfaces extending arbitrarily close to the event
horizon in any region of the spacetime.
The event horizon, then, is the boundary of the region containing outer
trapped surfaces and in this case Eardley's conjecture is indeed true.

\begin{figure}
      \begin{center}
    \resizebox{13cm}{!}{\includegraphics{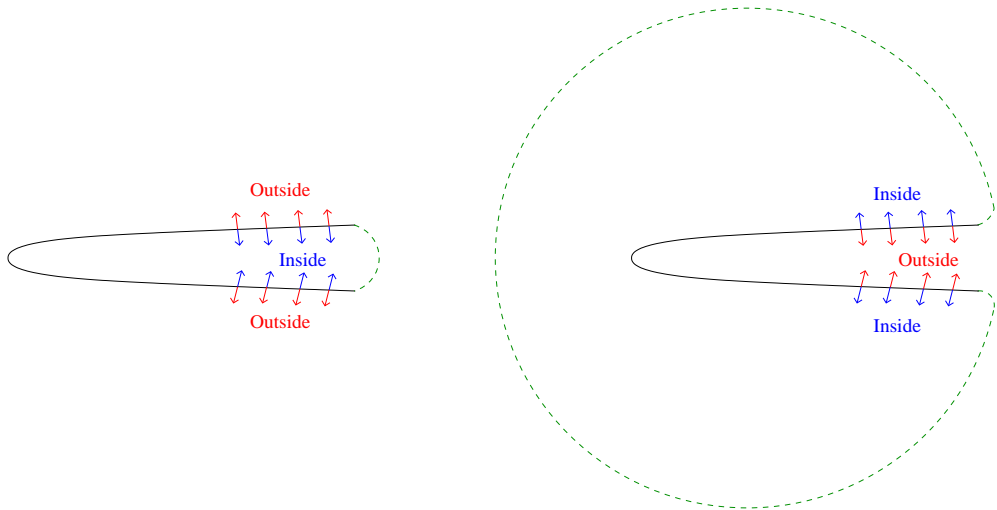}}
  \end{center}
  \caption{Left, a narrow tube shaped like a worm's skin. The ingoing and
outgoing directions are shown. Right, a similar tube closed off differently.
Now it is closed inside-out, like the face of Pacman. As a result the ingoing
and outgoing directions of the tube on the left, are interchanged. Therefore,
a portion of an inner trapped surface (the worm) can be used in constructing
an outer trapped surface (the Pacman).
The specific closing off inside-out will be done in the
far future, inside the apparent horizon, where 2-spheres are outer trapped.}
  \label{ClosingFig}
\end{figure}

This will be achieved by constructing an outer trapped surface.
Starting with the initial point that lies inside the event horizon, and
which might be located in a flat portion, a spacelike narrow tube
is constructed.
This tube stays inside the event horizon and corresponds to Eardley's tendril.
The tube reaches inside the apparent horizon in the
far future and it then tends to a region very close to the singularity.
The idea is to close the surface off in a specific way once it is close
to the singularity, thereby ensuring that the
resulting surface is outer trapped.
A sketch of this idea is shown in Fig.\ \ref{ClosingFig}.
Closing this tube into a surface shaped like a worm's skin, could lead to an
inner trapped surface as can be seen on the left.
Inner trapped surfaces certainly exist in flat space.
If instead the surface is closed off inside-out, like
the face of Pacman,\footnote{The surface that is closed off inside-out is
more accurately described as the boundary
of a ball with a cylinder removed.} then it may be possible
to obtain an outer trapped surface in this way, since this
procedure interchanges the ingoing and outgoing directions, as can be seen on
the right.
Recall that inside the spherically symmetric apparent horizon, 2-spheres are
outer trapped.
Since the closing off is done in the far future, inside the apparent horizon,
then it will be possible to keep the expansion negative in that region as well.

It will also be shown in this work that in Vaidya spacetimes, the
region containing trapped surfaces does not, in general, extend
everywhere to the event horizon.
As shown by Schnetter and Krishnan, the boundary of the region containing
trapped surfaces may extend to the flat region of a Vaidya spacetime.
However, the result obtained here shows that in general, this boundary is
separated from the event horizon in the flat region.


\section{Main result and key ideas} \label{mainresult}

Consider a Vaidya spacetime with a metric given by (\ref{Metric})
with $R(v)$ some non-negative, non-decreasing, smooth function of $v$ that is
bounded from above with a least upper bound, $R_0$.

Since $R(v)$ is bounded by $R_0$ then this spacetime is asymptotically flat and
tends to Schwarzschild spacetime in the far future. It follows that this
spacetime (for the non-trivial $R_0 \neq 0$ case) contains a spherically
symmetric event horizon, which can be described by the equation $r=r_{eh}(v)$.
The function $r_{eh}$ cannot be specified without full knowledge of the
function $R(v)$. However, like $R(v)$, this function is non-decreasing and in
addition $r_{eh}(v) \geq R(v)$ with equality holding at $v_0$ only if $R(v)$ is
constant for all $v \geq v_0$ (i.e.\ if the spacetime is exactly
Schwarzschild in the future).
If the spacetime contains a flat region, i.e.\ if there exists
some $v_\text{flat}$ such that $R(v)=0$ for all $v\leq v_\text{flat}$, then
the event horizon extends to this region as well.

Consider a point in this spacetime that lies inside the event horizon, but may
be arbitrarily close to it.
It lies, in the coordinates above, at $r_0$ and $v_0$ satisfying
$r_0 < r_{eh}(v_0)$ and, without loss of generality, this point is assumed to
lie at $\theta=0$.

The main result of this work is a proof by construction of the existence of an
outer trapped surface that contains this point. More precisely, the following
proposition is proven:

{\bf Proposition}: Given a Vaidya spacetime with metric as
in (\ref{Metric}) such that
$R(v)$ is a non-negative, non-decreasing, bounded function with $R_0>0$
the least upper bound of $R(v)$, and given any point that lies inside the
event horizon, then there exists a compact smooth spacelike 2-manifold, such
that the expansion, $\Theta$, of outgoing future-directed null geodesics
normal to it is everywhere negative.

This leads directly to the following result:

{\bf Corollary}: In any Vaidya spacetime of bounded total mass, the outer
boundary of the region containing outer trapped surfaces is the event horizon.

The proposition will be proved in the next section by direct construction.
Before doing so, it is worthwhile to understand the general situation and
the central ideas of the construction.

As discussed earlier, 2-spheres with $r<R(v)$ are outer trapped. Therefore,
the challenge is to construct an outer trapped surface containing
a point that is located at some given $r=r_0$ and $v=v_0$ such that
$R(v_0) < r_0 < r_{eh}(v_0)$.

Since the desired outer trapped surface cannot be spherically symmetric, the
next simplest option is an axisymmetric surface. The first central idea in the
present work is the particular way in which this surface is constructed.
A spacelike vector field, $w^a$, is defined.
This vector field is orthogonal to the axial Killing field, 
$c^a \equiv (\frac{\partial}{\partial \phi})^a$.
The integral curve of $w^a$ that contains the initial point is
chosen.
This curve is then translated by the axial Killing field, $c^a$,
and this results in a surface.
Since the Killing field has closed orbits then provided that the integral
curve starts and ends at the axis and does not intersect it in-between,
the resulting surface will be compact.
If $w^a$ is smooth then the surface obtained in this way is smooth
except possibly for the north and south poles.
Extra care will be taken at the poles to ensure the surface is smooth
everywhere.

Next, consider the expansion, $\Theta$. The outgoing, future directed, null
rays orthogonal to this surface are given by some null vector field $l^a$ such
that $l^a$ is orthogonal to the vector fields $w^a$ and $c^a$.
There still remains a scaling freedom for $l^a$.
However, this freedom does not affect the sign of the expansion, and since for
a surface to be outer trapped it is the sign of the expansion that matters,
then any convenient choice will do.

The expansion of $l^a$ is given by $\Theta=q^{ab} \nabla_a l_b$ where
$\nabla_a$ is the covariant derivative associated with $g_{ab}$ and
$q^{ab}$ is the inverse of the induced metric on the 2-surface given by
\bean
\label{InducedMetric}
q^{ab} = \frac{w^a w^b}{w^c w_c} + \frac{c^a c^b}{c^c c_c}
\eean

Given a choice of a spacelike vector field $w^a$ as above, such that the
expansion of outgoing null rays, $l^a$, is negative along the integral curve
of $w^a$ then the surface obtained in this way is, as desired,
an outer trapped surface.

There are other, more direct, ways of specifying a 2-surface in spacetime.
For example, two functions in spacetime could be defined such that the
intersection of two level surfaces of these functions is a 2-surface.
However, it is much more difficult to use such direct methods for all
Vaidya spacetimes with a metric as in (\ref{Metric}) and with any $R(v)$ that
satisfies the conditions above, since they require the global specification
of the surface ``all at once'' such that $\Theta<0$ everywhere.
Instead, using the method employed here, one defines a vector field in
spacetime.
This allows one to make appropriate ``local adjustments'' to the surface
in the process of defining it.
Thus, one is able to ensure that the expansion is negative by such ``local
adjustments''.

\begin{figure}
      \begin{center}
    \resizebox{12cm}{!}{\includegraphics{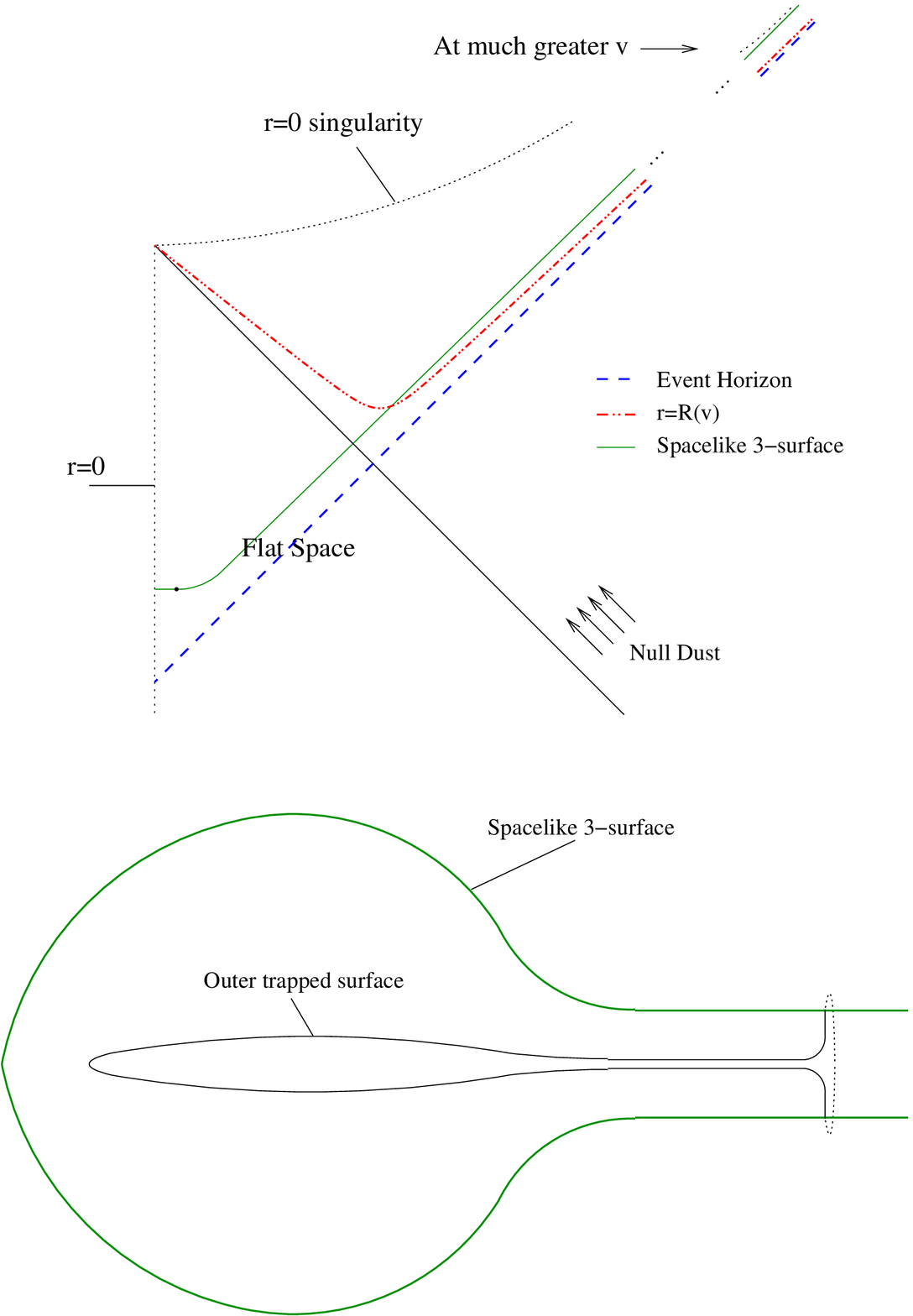}}
  \end{center}
  \caption{Top: A spacetime diagram of a Vaidya spacetime. Two angular
dimensions are suppressed. Points in this diagram are, therefore, 2-spheres.
A spherically symmetric spacelike 3-surface is shown. Bottom: A schematic
of this 3-surface. One angular dimension is not shown.
The desired outer trapped surface is the line traced on this 3-surface.}
  \label{SpacetimeFig}
\end{figure}

The central idea leading to a suitable choice of the vector field, $w^a$, is
that proximity to the axis combined with a specific closing of the 2-surface
will enable maintaining the expansion negative.
This was described earlier and was shown in Fig.\ \ref{ClosingFig}.
Fig.\ \ref{SpacetimeFig} shows how this idea is actually implemented in a
Vaidya spacetime.
The top diagram is a spacetime diagram of the collapse of null dust in
flat space.
In this diagram a spherically symmetric spacelike slice that contains the
desired outer trapped surface is shown.
The bottom diagram depicts this spacelike 3-surface.
The outer trapped surface including its angular ($\theta$) dependence
is shown as a line in this 3-surface. 
As can be seen in the bottom diagram of Fig.\ \ref{SpacetimeFig}, the
outer trapped surface remains close to the axis for the most part, i.e.\ at a
very small $\theta$.
Once the conditions are right, the 2-surface closes off in the particular way
described earlier, as can be seen at the right side of the bottom diagram.

A sketch of the construction is now given. Full details are given in the
next section.
In what follows the components of the vector fields are given with
respect to the coordinates $v,r$, and $\theta$ used in the metric
above.\footnote{There are no expressions containing $\phi$, since,
as discussed earlier, $w^a$ is orthogonal to the axial Killing vector.}

The vector field $w^a$ is constructed by a sequence of smooth transitions
between five vector fields.
Each one of these is used only in a restricted region of the spacetime.
This is achieved by taking $w^a$ to be of the form:
\bean
\label{wgeneral}
w^a &=& (1-\alpha_{1,2}) w_1^{\ a} + \alpha_{1,2}(1-\alpha_{2,3}) w_2^{\ a}
+ \alpha_{2,3}(1-\alpha_{3,4}) w_3^{\ a} \nonumber \\
&+& \alpha_{3,4}\,\cos\alpha_{4,5}\, w_4^{\ a}
+ \sin\alpha_{4,5}\, w_5^{\ a}
\eean
and the five vector fields are given by
\begin{align}
w_1^{\ a} &= \ 2R_0\,r\sin\theta \Big(\frac{\partial}{\partial v}\Big)^a +
2R_0 \Big(r-\frac{R(v)}{2}\Big) \sin\theta
\Big(\frac{\partial}{\partial r}\Big)^a & \nonumber \\
&+ \Big(2R_0 \cos\theta\ -r\Big) \Big(\frac{\partial}{\partial \theta}\Big)^a
& \label{w1} \\[5pt]
w_2^{\ a}&= 2 \Big(\frac{\partial}{\partial v}\Big)^a & \nonumber \\
&+ \Big(1-\frac{R(v)}{r} + 
\frac{\sqrt{4 R_0^{\ 2}-4 R_0\,r\cos\theta + r^2} - (2R_0 \cos\theta\,-r)}
{\sqrt{4R_0^{\ 2}-4R_0\,r\cos\theta + r^2} + (2R_0\cos\theta\,-r)} \Big)
\Big(\frac{\partial}{\partial r}\Big)^a
& \label{w2} \\[5pt]
w_3^{\ a} &= 2 \Big(\frac{\partial}{\partial v}\Big)^a +
\Big(1-\frac{R(v)}{r}(1-\frac{\eps}{r})\Big)
\Big(\frac{\partial}{\partial r}\Big)^a\
& \label{w3} \\[5pt]
w_4^{\ a} &= 2\Big(\frac{\partial}{\partial v}\Big)^a
& \label{w4} \\[5pt]
w_5^{\ a} &= \frac{2}{r}\,\sqrt{1-\frac{R(v)}{r}}\, 
\Big(\frac{\partial}{\partial \theta}\Big)^a
& \label{w5}
\end{align}
where $\eps$ is some positive constant to be specified later.

The functions $\alpha_{i,i+1}$ appearing in (\ref{wgeneral}), are referred to
as transition functions, i.e.\ these are functions that facilitate the
transition from one vector field to the next.
Except for the last transition function, $\alpha_{4,5}$, the functions
$\alpha_{i,i+1}$ are all smooth, non-decreasing functions of $v$ into $[0,1]$,
and are chosen so that in a neighborhood where one such $\alpha$ is not
identically $0$ or $1$ then all the other $\alpha$s are
fixed to either $0$ or $1$ identically there.\footnote{In the case of the
last transition, which, for convenience, uses $\cos\alpha$ and $\sin\alpha$,
instead of $1-\alpha$ and $\alpha$, respectively, the transition
function $\alpha_{4,5}$ is instead a smooth function
into $[0,\frac{\pi}{2}]$. As a result, in regions of the spacetime where
$\alpha_{4,5}$ is not identically $0$ or $\frac{\pi}{2}$, the other $\alpha$s
are fixed to either $0$ or $1$.}
This can be explained more clearly as follows: The smooth
functions $\alpha_{i,i+1}$ result in the division of the spacetime into
several disjoint regions that border each other.
First, there is a region where $w^a=w_1^{\ a}$.
This region contains the initial point.
Next, there is a transition region where $w^a$ is a smooth linear combination
of $w_1^{\ a}$ and $w_2^{\ a}$.
The next region is such that $w^a=w_2^{\ a}$ there.
This is followed by another transition region, and so on.
The integral curve that starts at the initial point will be shown to traverse
all these different regions.

Here is a rough description of what this surface is like.
Initially the surface starts at $\theta=0$ and $\theta$ increases to some
small fixed value.
Meanwhile, the surface continues towards increasing $v$ and $r$, tending closer
to the event horizon.
In the top diagram of Fig.\ \ref{SpacetimeFig}, this is the horizontal part of
the 3-surface in a small neighborhood of the initial point.
Next, the surface starts following the event horizon into the future at
constant $\theta$, always remaining inside the event horizon.
At late $v$ -- when the 3-surface $r=R(v)$ tends to the event horizon as the
spacetime asymptotes to Schwarzschild -- the constructed surface is finally
inside the 3-surface $r=R(v)$, as can be seen in the top diagram of
Fig.\ \ref{SpacetimeFig}.
At this stage $r$ starts decreasing.
This is shown in the bottom diagram of Fig.\ \ref{SpacetimeFig}.
The surface, next, settles to very small $r$, close to the black hole
singularity.
This can be seen in the top diagram of Fig.\ \ref{SpacetimeFig}, at much
larger $v$.
In the bottom diagram, this is the part where the geometry tends to a
cylinder with a constant thickness.
Finally, the surface closes off like a 2-sphere, following the idea
described earlier, as shown in the bottom diagram of Fig.\ \ref{SpacetimeFig}.


\section{Details}

The proposition is now proved by presenting the full details of the
construction.

Given a metric of the form (\ref{Metric}) with $R(v)$ satisfying the
conditions above, consider an event, i.e.\ a point in this spacetime,
which without loss of generality can be taken to lie
at the north pole, $\theta=0$, at some $v=v_0$ and $r=r_0$ such that
$R(v_0) < r_0 < r_{eh}(v_0)$. The first inequality here ensures that
finding an outer trapped surface is non-trivial, since if $r_0$ were smaller
than $R(v_0)$ then the 2-sphere with constant $r$ and $v$ passing through the
point would be outer trapped.
The second inequality follows since the point of interest lies strictly
inside the event horizon (though perhaps arbitrarily close to it).

The following notation is used. Let $v_i,r_i$, and $\theta_i$
stand for the value of the coordinates $v,r$, and $\theta$ respectively, when
the transition from $w_i^{\ a}$ to $w_{i+1}^{\ \ \ a}$ starts.
Let $v_{i,i+1},r_{i,i+1}$, and $\theta_{i,i+1}$ be the values of these
coordinates at the end of the transition to $w_{i+1}^{\ \ \ a}$.

Given the value of $w^a$ in some region, one needs an expression for $l^a$ in
order to evaluate the expansion. Since $l^a$ can be rescaled by an arbitrary
function, there remains some freedom in choosing the components of $l^a$. Here,
a choice will be made by fixing\footnote{This is possible
since the outgoing null vector field satisfies
$l^a\Big(\frac{\partial}{\partial r}\Big)_a > 0$. As a result this component
is always positive and can be chosen to be equal to two.}
the component of $l^a$ along $(\frac{\partial}{\partial v})^a$ to be two.
The components of $l^a$ along $(\frac{\partial}{\partial r})^a$ and
$(\frac{\partial}{\partial \theta})^a$ can then be uniquely determined as
follows. The two conditions $l^a l_a = 0$ and $l^a w_a = 0$ translate into
equations that when solved uniquely determine the component along
$(\frac{\partial}{\partial r})^a$. These also yield two possible solutions
for $(\frac{\partial}{\partial \theta})^a$ since the conditions used so far
allow $l^a$ to be either outgoing or ingoing. The outgoing direction, as
discussed earlier and shown in Fig.\ \ref{SpacetimeFig} is the solution tending
towards smaller $\theta$, i.e.\ the more negative solution.
This procedure is followed for the entire sequence of transitions including
during the transitions themselves.

Using (\ref{InducedMetric}), the expansion, $\Theta$, can be written as:
\bean
\label{Expansion}
\Theta(w) &=& q^{ab}\nabla_a l_b \nonumber \\
&=& \Big(\frac{w^a w^b}{w^c w_c} +
\frac{c^a c^b}{c^c c_c} \Big) \nabla_a l_b \nonumber \\
&=& -\frac{w^a l^b \nabla_a w_b}{w^c w_c} +
\frac{c^a l^b \nabla_b c_a}{c^c c_c} \nonumber \\
&=& -\frac{w^a l^b \nabla_a w_b}{w^c w_c} +
\frac{1}{2} l^b \nabla_b \log (c^a c_a)
\eean
where in the third equality, $l^a w_a=0$ was used in getting the first term and
$l^a c_a=0$ and $c^a$ being a Killing vector field were used in getting the
second term.
Throughout this work the expansion is evaluated using this expression.

Getting an outer trapped surface requires precise control over the location
of the integral curve.
This is achieved with a suitable tuning of the transition functions
$\alpha_{i,i+1}$ that control the five vector fields and basically turn
the vector fields on and off in certain regions of the spacetime.

Except for the final transition, from $w_4^{\ a}$ to $w_5^{\ a}$, all the
transitions are taken to be of the form
\bean
\label{Transition}
w^a = \Big(1-\alpha_{i,i+1}(v)\Big) w_i^{\ a} + \alpha_{i,i+1}(v)
w_{i+1}^{\ \ \ a}
\eean
As functions of $v$ this means that $w^a$, except for the final transition,
simply changes with $v$ as follows.
At $v\leq v_1$, $w^a=w_1^{\ a}$.
A transition via $\alpha_{1,2}$ takes place and at any intermediate $v$, i.e.\ 
$v_1<v<v_{1,2}$, the vector field $w^a$ is some linear combination of
$w_1^{\ a}$ and $w_2^{\ a}$.
At $v_{1,2} \leq v \leq v_2$, $w^a=w_2^{\ a}$, and so on.

For simplicity, the form of the transition is picked once and is used
repeatedly.
Pick a smooth function $f(x)$ such that $f(x)$ is non-decreasing,
$f(x) = 0$ for $x\leq 0$, $f(x)=1$ for $x\geq 1$, and finally
$f'(x)>0 \ \forall x\in(0,1)$.
Given this choice\footnote{Here is one possible choice. Let $g(x)$ be the
smooth function defined by $g(x)=e^{-1/x^2}$ for $x>0$, and $g(x)=0$
otherwise.
It follows that $f(x)=e^{-g(1-x)/x^2}$ for $x>0$, and $f(x)=0$ otherwise,
satisfies the desired properties.}\,of $f(x)$, all of the transition
functions, except the last one, will be of the form
\bean
\label{GeneralTransition}
\alpha_{i,i+1}(v) = f\Big(\lambda (v-v_i)\Big)
\eean
with $\lambda = \frac{1}{v_{i,i+1} - v_i}$.
It follows that $\alpha_{i,i+1}(v)=0$ for $v \leq v_i$ and that
$\alpha_{i,i+1}(v)=1$ for $v \geq v_{i,i+1}$.

As a result of using this canonical transition, it remains only to choose
the various starting and ending regions for the transitions.
In each transition it will be shown that with the choices made, the integral
curve along $w^a$ reaches the desired regions and the expansion along the
integral curve of $w^a$ is always negative.

Two parameters will be used in the choosing of the beginning and ending
regions for the transition functions.
These are given by
\bean
\eps &=& \min \Big(\frac{r_{eh}(v_0)-r_0}{7},\,\eps^*,\,
\frac{R_0}{16}\Big)
\label{parmeps} \\ \nonumber \\
\tilde\theta &=& \min \Big(\frac{\sqrt{\eps\, R_+}}{2 R_0} ,\,
\sqrt{\frac{\eps}{v^*_{r_0} -v_0}},\,
\theta^*_1,\,\theta^*_2,\,\theta^*_3,\, \frac{\pi}{8}\Big)
\label{parmtheta}
\eean
where $v^*_{r_0}$ satisfies $R(v^*_{r_0})=r_0$, and $R_+$ is given by
\bean
\label{R+}
R_+ &=& \left\{ \begin{array}{l@{\quad\quad\quad}l}
\frac{1}{2}R(v_0), & R(v_0)>0 \vspace{5pt}\\
\frac{1}{2}r_0, & R(v_0)=0 \\
\end{array} \right.
\label{parmRplus}
\eean
The parameters $\theta^*_1,\theta^*_2,\theta^*_3 \text{, and }\eps^*$ are
chosen in the following sections.
The choice of $\theta^*_1$ is made at the end of Section \ref{details1}.
$\theta^*_2$ is chosen in Section \ref{details2}.
The choice of $\theta^*_3$ is made at the beginning of Section \ref{details3}.
Finally, $\eps^*$ is chosen in Section \ref{details5}.
It is important to note that the choices of
$\theta^*_1,\,\theta^*_2,\,\theta^*_3,\text{and }\eps^*$ are all independent
of each other.\footnote{If this were not the case, then a choice of one of
these later in the construction, could affect, via (\ref{parmtheta}), a choice
that has already been made. Thus, this independence is important to avoid any
circular logic in making these choices.}
The specification of
$\theta^*_1,\,\theta^*_2,\,\theta^*_3,\text{and }\eps^*$ is discussed later
purely for convenience, so that the relevant conditions for making these
specifications will be at hand.

Since $\theta^*_1,\,\theta^*_2,\,\theta^*_3,\text{and }\eps^*$ are all positive
then it follows that $\tilde\theta$ and $\eps$ are positive as well.
The reasons for these choices will be clear once the parameters are used in
the construction.

Along the integral curve $r$ can be considered as a function of $v$ and
therefore it is possible to define the following function:
\bean
\label{tilder}
\tilde r(v) \equiv r(v)-r_{eh}(v)
\eean
This function serves to keep track of where is the integral curve
relative to the event horizon, for a given $v$.

\subsection{Initially $w^a=w_1^{\ a}$ : Starting at the north pole}
\label{details1}

At the initial point, at the north pole, the construction starts with
$w^a=w_1^{\ a}$ where $w_1^{\ a}$ was given\footnote{The choice of
$w_1^{\ a}$ was originally motivated by considering a 2-sphere in flat space
and shifting it by $2R_0$ along the $z$-axis.}\  by (\ref{w1}).

As mentioned earlier, the north pole, $\theta=0$, is one point where extra
care must be taken to ensure the surface is smooth.
This choice of $w_1^{\ a}$ guarantees that the resulting surface will be
smooth at the north pole as follows.
The integral curve of $w_1^{\ a}$ near the north pole can be
parametrized by $\mu=\cos\theta$. This gives a system of ordinary
differential equations for $v$ and $r$ as functions of $\mu$ with smooth
coefficients in a neighborhood of the north pole. It follows that
in a neighborhood of the north pole, in the resulting surface, $v$ and $r$ are
smooth functions of $\mu$. Changing into Cartesian coordinates one can then
verify that the surface is smooth at the north pole.

The outgoing null vector field, $l^a$, is needed for evaluating the expansion.
Following the procedure discussed earlier, it is found to be
\bean
\label{l1}
l_1^{\ a} \ = && 2 \Big(\frac{\partial}{\partial v}\Big)^a +
\frac{2}{1+a(r,\theta)}\Big(1-\frac{R(v)}{2r}- a(r,\theta)\frac{R(v)}{2r}
\Big) \Big(\frac{\partial}{\partial r}\Big)^a \nonumber \\
&& -\ \frac{2 a(r,\theta)}{1+a(r,\theta)} \, \frac{2R_0 \sin\theta}
{r \sqrt{4R_0^{\ 2}-4R_0\,r\cos\theta + r^2}}
\Big(\frac{\partial}{\partial \theta}\Big)^a
\eean
with the function $a(r,\theta)$ given by
\bean
\label{funca}
a(r,\theta) = \frac{\sqrt{4R_0^{\ 2}-4R_0\,r\cos\theta + r^2}}
{2R_0 \cos\theta -r}
\eean
Using (\ref{Expansion}) the expansion is found to be
\bean
\label{exp1short}
\Theta(w_1)\, &=& \, -\frac{2\Big(r^2+(2R_0-r)R(v)\Big)}{r^2(2R_0-r)}
+ O(\theta^2)
\nonumber \\
&<& -\frac{1}{R_0}+ O(\theta^2)
\eean
where in the last inequality, $R(v)<r<R_0$ was used.
This holds when $w^a=w_1^{\ a}$.

It follows that there exists some $\theta^*_1>0$ such that
if $\theta\leq\theta^*_1$ then the expansion is negative.
Thus $\theta^*_1$ is now chosen.
As will be seen below, when $w^a=w_1^{\ a}$, $R(v)<r<R_0$ and
$\theta\leq\frac{1}{2}\tilde\theta\leq\frac{1}{2}\theta^*_1$ will be
maintained.
As a result, during this stage the expansion is negative.

\subsection{Transition from $w_1^{\ a}$ to $w_2^{\ a}$ : The integral curve
becomes radial} \label{details2}

Initially $R(v)=R(v_0)$, $\tilde r(v)\leq -7\eps$, and $\theta=0$.
In order to keep control of where the integral curve eventually reaches,
it is desired to have the transition to $w_2^{\ a}$ start when
\bean
R(v)\leq R(v_0)+\delta,\quad \tilde r(v) \leq -6\eps,\quad
\text{and}\quad 0<\theta\leq\frac{1}{2}\tilde\theta
\eean
where $\delta>0$ is given by
\bean
\label{delta}
\delta = \frac{1}{4}\Big(r_0 - R(v_0)\Big)
\eean

Since $w_1^{\ \theta}\equiv w_1^{\ a}\,d\theta_a$ is positive and bounded away
from zero for $\theta\leq\frac{\pi}{16}$ and since, from (\ref{w1}), the
integral curve stays away from the origin, $r=0$, and its components are
all bounded, then the integral curve along $w_1^{\ a}$ reaches
$\theta=\frac{\pi}{16}$.

Let $v_1$ be the smallest value of $v>v_0$ such that one of the following
occurs along the integral curve of $w_1^{\ a}$ starting at the
initial point: (i) $R(v)=R(v_0)+\delta$, (ii) $\tilde r(v)=-6\eps$, or
(iii) $\theta=\frac{1}{2}\tilde\theta$.
This ensures that $v_1$ is reached and the transition to $w_2^{\ a}$
starts.
The choice of $v_1$ implies that $r_1 - R(v_1) \geq 3\delta$, 
$\tilde r(v_1) \leq -6\eps$ and $\theta_1 \leq \frac{1}{2}\tilde\theta$.

As previously indicated, the transition is taken to be of the form
\bean
\label{transition12}
w^a = \big(1-\alpha_{1,2}(v)\big) w_1^{\ a} + \alpha_{1,2}(v) w_2^{\ a}
\eean
where $w_2^{\ a}$ was given in (\ref{w2}).
The transition function $\alpha_{1,2}$ is given by (\ref{GeneralTransition})
with $v_1$ as chosen above and $v_{1,2}$, the value of $v$ where the 
transition ends, is chosen next.

Let $v_{1,2} =v_1 + \eps_{1,2}$ where $\eps_{1,2}>0$ is now chosen.
Since $R(v_1)\leq R(v_0)+\delta$ then it follows that a sufficiently small
$\eps_{1,2}$ implies that $R(v_{1,2})\leq R(v_0)+2\delta$.
Therefore, this is one upper bound on $\eps_{1,2}$ so that if 
$\eps_{1,2}$ is chosen smaller than this bound, then $R(v)$ is kept under
control.

From the form of the transition, (\ref{transition12}), and since
it starts at $\theta_1>0$ then it can easily be verified that
$\frac{d\theta}{dv}$ and $\frac{dr}{dv}$ are both non-negative and bounded
from above during the entire transition (i.e.\ $\forall\alpha_{1,2}\in[0,1]$).

As a result, a sufficiently small $\eps_{1,2}>0$ implies that
$r_{1,2}-r_1\leq\eps$ and that
$\theta_{1,2}-\theta_1\leq\frac{\tilde\theta}{2}$.
Now, $r_{1,2}-r_1\leq\eps$ implies that 
$\tilde r(v_{1,2})-\tilde r(v_1)\leq\eps$ as well, since recall that
$\tilde r(v)\equiv r(v) - r_{eh}(v)$ and $r_{eh}(v)$ is non-decreasing.

Therefore, a sufficiently small $\eps_{1,2}$ is chosen such that
during the transition, including when it ends, $R(v)\leq R(v_0)+2\delta$,
$\tilde r(v)\leq -5\eps$, and $\theta\leq\tilde\theta$.

It is important to verify that given $\alpha_{1,2}(v)$, the integral curve
along $w^a$ that is given by (\ref{transition12}) actually reaches\
$v=v_{1,2}$, i.e.\ verify that the transition ends and reaches the region in
spacetime where $w^a=w_2^{\ a}$.
Since during the transition, the component $w^v\equiv w^a\,dv_a$ is positive
and bounded away from zero, and since the other components are bounded and
the integral curve does not tend to $r=0$, then the integral curve reaches
$v=v_{1,2}$.

Consider, next, the expansion throughout the transition.
It is helpful to note that $l_2^{\ a}=l_1^{\ a}$, i.e.\ $w_2^{\ a}$ has the
same null field orthogonal to it, as $w_1^{\ a}$ does.
This follows since one can write
\bean
\label{w1andw2}
w_1^{\ a} = C_w\, w_2^{\ a} + C_l\, l_1^{\ a}
\eean
where $C_w$ and $C_l$ are some smooth functions and in the region of interest
$C_w$ is positive while $C_l$ is negative.

Since $l_1^{\ a}=l_2^{\ a}$ then it follows that throughout the
transition, the expansion does not depend on $\alpha_{1,2}'(v)$.
Combining (\ref{transition12}) and (\ref{w1andw2}) one obtains
\bean
\label{newtransition12}
w^a = f w_2^{\ a} + g l_1^{\ a}
\eean
where $f=(1-\alpha_{1,2})C_w + \alpha_{1,2}$ and $g=(1-\alpha_{1,2})C_l$.

Using (\ref{Expansion}) the expansion is then found to be given by
\bean
\label{exp12}
\Theta\Big((1-\alpha_{1,2})w_1+\alpha_{1,2}w_2\Big) &=& \Theta(w_2)
\nonumber \\
&-& \frac{(1-\alpha_{1,2})C_l}{(1-\alpha_{1,2})C_w+\alpha_{1,2}}
\, \frac{l_1^{\ a} l_1^{\ b} \nabla_a w_{2b}}{w_2^{\ c} w_{2c}} 
\eean

Consider the first term in the expansion above, the expansion of $w_2^{\ a}$.
Using $l_2^{\ a}=l_1^{\ a}$ and (\ref{Expansion}) this is found to be
\bean
\label{exp2}
\Theta(w_2) &=&
\frac{\Big(R(v)-4\,r\Big) 
 + \frac{R(v)\big(2R_0\cos\theta-r\big)}
{\sqrt{4R_0^{\ 2}-4R_0\,r\cos\theta + r^2}} }
{r \Big(2R_0\cos\theta-r +
\sqrt{4R_0^{\ 2}-4R_0\,r\cos\theta + r^2} \Big)} \nonumber \\
&=& -\frac{2r-R(v)}{r(2R_0-r)} + O(\theta^2)
\eean
There exists some $\theta^*_2>0$ such that $\Theta(w_2)$ is
negative for all $\theta\leq\theta^*_2$ when $R(v)<r<r_{eh}(v)$.
This $\theta^*_2$ is chosen and appears in (\ref{parmtheta}).

Consider, next, the second term in (\ref{exp12}).
The dependence on $\alpha_{1,2}$ is given by
\bean
\label{alpha12term}
\frac{(1-\alpha_{1,2})C_l}{(1-\alpha_{1,2})C_w+\alpha_{1,2}}
\eean
Since in the region of interest $C_w$ is positive and $C_l$ does not change
sign, then it follows that this expression is monotonic in $\alpha_{1,2}$.
As a result, the maximum for (\ref{exp12}) is attained at $\alpha_{1,2}=0$
or $\alpha_{1,2}=1$.
When $\alpha_{1,2}=0$, (\ref{exp12}) is the expansion of $w_1^{\ a}$ and when
$\alpha_{1,2}=1$, it is the expansion of $w_2^{\ a}$.
As a result it follows that
\bean
\label{exp12smallerexp2}
\Theta((1-\alpha_{1,2})w_1+\alpha_{1,2}w_2)) \, \leq
\,\max\Big(\Theta(w_1),\Theta(w_2)\Big)
\eean
Since the expansions of $w_1^{\ a}$ and $w_2^{\ a}$ are both negative
then the expansion during the transition is negative as well.

In the region where $w^a=w_2^{\ a}$, it will be seen below that
$\theta\leq\theta^*_2$ and that $R(v)<r<r_{eh}(v)$. 
Hence, the expansion along the integral curve of $w^a$ is negative in
the region where $w^a=w_2^{\ a}$.

\subsection{Transition from $w_2^{\ a}$ to $w_3^{\ a}$ : Once $R$ is positive}
\label{details3}

Consider, first, the value $v_2$ of $v$, at which the transition begins.
Let $v_2$ be the smallest value to satisfy $R(v_2) = R(v_0)+3\delta$,
where $\delta$ was given by (\ref{delta}).
This means that when the transition to $w_3^{\ a}$ starts, $R(v)$ is
already positive.
This will turn out to be crucial for keeping the expansion during the
transition negative.

Along the integral curve of $w_2^{\ a}$, for all $v\leq v_2$, it will now
be shown that $\tilde r(v) \leq -4\eps$.
The condition set forth for $R(v_2)$ already takes care of controlling the
increase in $R(v)$.
Since $\theta$ is constant, its increase is certainly under control.

Consider the following function that is given in (\ref{w2}) as part of
a component of $w_2^{\ a}$.
\bean
\label{funch}
h(r,\theta)&=&\frac{\sqrt{4R_0^{\ 2}-4R_0\,r\cos\theta + r^2} -
(2R_0\cos\theta\,-r)}
{\sqrt{4R_0^{\ 2}-4R_0\,r\cos\theta + r^2} +
(2R_0\cos\theta\,-r)} \nonumber \\
&=& \frac{4R_0^{\ 2}\,{\theta }^2}{4(2R_0 - r)^2} + O(\theta^4) \,
< \, \theta^2 + O(\theta^4)
\eean
where for the last inequality, $r<r_{eh}(v)\leq R_0$ was used.
It follows that there exists some $\theta^*_3>0$ such that for all
$\theta\leq\theta^*_3$ and when $r<r_{eh}(v)$ then
\bean
\label{hbound}
h(r,\theta)\leq 2 \theta^2
\eean
Such a $\theta^*_3$ is now chosen and since along the integral curve of
$w_2^{\ a}$, $\theta\leq\tilde\theta$ and $r<r_{eh}(v)$, then (\ref{hbound})
is satisfied there.

A bound on the change in $\tilde r(v)\equiv r(v)-r_{eh}(v)$ is obtained
as follows.
\bean
\label{dtilder}
\frac{d \tilde r(v)}{dv} &=& \frac{1}{2}
\Bigg(-\frac{R(v)}{r} +
\frac{\sqrt{4R_0^{\ 2}-4R_0\,r\cos\theta + r^2} - (2R_0\cos\theta\,-r)}
{\sqrt{4R_0^{\ 2}-4R_0\,r\cos\theta + r^2}
+ (2R_0\cos\theta\,-r)} \nonumber \\
&+& \frac{R(v)}{r_{eh}(v)} \Bigg) 
\  \leq \  \frac{1}{2} h(r,\theta) \  \leq \  \theta^2
\eean
where, for the first equality,
$dr_{eh}/dv=\frac{1}{2}\Big(1-\frac{R(v)}{r}\Big)$ was used and
$dr/dv$ was obtained from (\ref{w2}).
The last inequality follows from (\ref{hbound}).

Given this bound it follows that
\bean
\label{deltatilder}
\tilde r(v_2) - \tilde r(v_{1,2}) \leq \theta^2 (v_2-v_{1,2})
\leq \theta^2 (v^*_{r_0}-v_0) \leq \eps
\eean
where recall that $v^*_{r_0}$ satisfies $R(v^*_{r_0})=r_0$ and
therefore $v^*_{r_0}>v_2$.
For the last inequality, it follows from (\ref{parmtheta}) that
$\theta\leq \tilde\theta\leq\sqrt{\frac{\eps}{v^*_{r_0} -v_0}}$.

At $v=v_{1,2}$, when $w^a=w_2^{\ a}$ initially, $\tilde r \leq -5\eps$ holds.
It follows from (\ref{deltatilder}) that at $v=v_2$, $r \leq r_{eh}(v) -4\eps$.

Consider next, the choice of $v_{2,3}$ where the transition ends.
Let $v_{2,3}=v_2 + \eps_{2,3}$ where $\eps_{2,3}>0$ is now chosen.
Once again, it is of interest to keep the changes in $\tilde r$ and $R(v)$
under control as well as make sure the expansion is negative.
This will be achieved by imposing three upper bounds on $\eps_{2,3}$

The derivative, $\frac{d\tilde r}{dv}$ is bounded from above throughout the
transition (i.e.\ $\forall\alpha_{2,3}\in[0,1]$) and hence if $\eps_{2,3}$ is
sufficiently small then $\tilde r(v_{2,3})-\tilde r(v_2)\leq\eps$.
This is the first upper bound on $\eps_{2,3}$.

Taking $\eps_{2,3}\leq v^*_{r_0}-v_2$ guarantees that by the end of
this transition $v_{2,3}\leq v^*_{r_0}$ and therefore $R(v_{2,3})\leq 
R(v^*_{r_0})=r_0$.
This is the second upper bound on $\eps_{2,3}$.
The third upper bound on $\eps_{2,3}$ will be specified after
discussing the expansion during the transition.

The vector field is transformed into $w_3^{\ a}$ via a transition of the
usual form,\footnote{Recall that by now $\alpha_{1,2}=1$ and in this region
all other transition functions $\alpha_{i,i+1}$ are zero, except for
$\alpha_{2,3}$. Hence locally the transition involves
only $w_2^{\ a}$ and $w_3^{\ a}$ in this manner.}
\bean
\label{transition23}
w^a = \Big(1-\alpha_{2,3}(v)\Big)w_2^{\ a} + \alpha_{2,3}(v) \,w_3^{\ a}
\eean
where $w_3^{\ a}$ was given in (\ref{w3}).

The vector field $l^a$ for this transition can easily be found, and
using (\ref{Expansion}), the expansion, $\Theta$, is found to be
\bean
\label{exp23}
&& \Theta\Big(\big(1-\alpha_{2,3}(v)\big) w_2+\alpha_{2,3}(v) w_3 \Big)
\nonumber \\
&& \quad\quad = \frac{1}{X} \Big(\frac{r^2}{R(v)}h(r,\theta) -\eps\Big)
\, \alpha_{2,3}'(v) \nonumber \\
&& \quad\quad -\ \frac{\alpha_{2,3}(v)\,\eps}{X}\,\frac{R'(v)}{R(v)}
\nonumber \\
&& \quad\quad
-\ \frac{\big(1-\alpha_{2,3}(v)\big)\big(X + r(\frac{r}{R(v)}-1)\big)}{2X}
\frac{\partial h(r,\theta)}{\partial r}
\nonumber\\
&& \quad\quad
-\frac{2}{r}\sqrt{\frac{R(v)}{r^2} X} \cot\theta
\nonumber \\
&& \quad\quad
+ \frac{2-\big(1-\alpha_{2,3}(v)\big) h(r,\theta)}{r} -\frac{R(v)}{r^2}
\nonumber \\
&& \quad\quad
-\frac{\big(1-\alpha_{2,3}(v)\big) h(r,\theta)\Big(\frac{r}{R(v)}-1\Big)}{X}
\eean
where $h(r,\theta)$ was given in (\ref{funch}) and
$X=\frac{r^2}{R(v)}\big(1-\alpha_{2,3}(v)\big) h(r,\theta) +
\alpha_{2,3}(v)\,\eps \,>\, 0$.
This expression consists of five terms.

The first term goes like $\alpha_{2,3}'(v)$ (second line).
This term is negative since
\bean
\frac{r^2}{R(v)}h(r,\theta)\leq \frac{2 \theta^2\, r^2}{R(v)} \leq
\frac{\eps}{2}\frac{R_+}{R(v)}\frac{r^2}{R_0^{\ 2}}<\frac{\eps}{2}
\eean
where (\ref{hbound}) was used in the first inequality and
(\ref{parmtheta}) was used in the second one.
The last inequality follows since, during this transition,
$r<R_0$ and $R_+<R(v)$, where $R_+$ was defined in (\ref{R+}).
Note that the coefficient in front of $\alpha'_{2,3}(v)$ is negative
and bounded away from zero for all $\alpha_{2,3}\in [0,1]$.
This fact will be used shortly.

The second term, the one that is proportional to $R'(v)$ (third line) is
always non-positive.
The third term is proportional to
$\frac{\partial h(r,\theta)}{\partial r}>0$ (fourth line).
This term is non-positive when $r>R(v)$ and this is satisfied
since $R(v) \leq R(v_0) + 4\delta = r_0 <r$ holds throughout the transition.

Finally there are two more terms, a term that goes like $\cot\theta$
(fifth line) and the remaining term, the last two lines.
These two combined, i.e.\ the last three lines, will only be considered
in the limits $\alpha_{2,3}(v)=0$ and $\alpha_{2,3}(v)=1$, since, it will be
shown that as long as this combined term is negative in both limits,
then a suitable choice of $\eps_{2,3}$, where the transition ends,
will guarantee that the last three lines can never make the expansion
non-negative.

When $\alpha_{2,3}(v)=0$, i.e.\ the beginning of the
transition, the combined term (last three lines) is negative.
When $\alpha_{2,3}(v)=1$, i.e.\ when the transition ends, this term is
negative provided that $\tan\theta < \frac{\sqrt{\eps R(v_{2,3})}}{r}$.
This is indeed satisfied, since
\bean
\label{tantheta}
\tan\theta < 2\theta \leq \frac{\sqrt{\eps R_+}}{R_0}
< \frac{\sqrt{\eps R(v_{2,3})}}{r}
\eean
where the first inequality certainly holds for $\theta\leq\frac{\pi}{8}$.
The next inequality follows from (\ref{parmtheta}).
Using $R(v_{2,3})\geq R(v_2)>R_{+}$ and $r<R_0$, the final inequality
immediately follows.

Since the combined term is negative for $\alpha_{2,3}=0$ and
$\alpha_{2,3}=1$ for all $v,r$, and $\theta$ satisfying the other conditions
above, then there exists some $\sigma>0$ such that the combined
term is negative for $\alpha_{2,3}<\sigma$ and for
$\alpha_{2,3}> 1-\sigma$.
It will now be shown that a suitable choice of $\eps_{2,3}$ can keep the
combined term negative for all $\alpha_{2,3}\in [0,1]$.

From the form of $\alpha_{2,3}$ in terms of $f(x)$, (\ref{GeneralTransition}),
it follows that
\bean
\label{alprime}
\alpha_{2,3}'(v) = \lambda\,f'(x)|_{x=\lambda(v-v_2)}
\eean
with $\lambda=(v_{2,3}-v_2)^{-1}=\eps_{2,3}^{\ \ -1}$.
Since $f'(x)>0$ for all $x\in(0,1)$ then there exists an inverse function
$f$ and therefore $\alpha_{2,3} \in [\sigma,1-\sigma]$ corresponds to $f(x)$
for $x\in[f^{-1}(\sigma),f^{-1}(1-\sigma)]$.
This closed interval for $x$ is compact and $f'(x)>0$ there.
It follows that $f'(x)$ has a minimum in this interval, $f'_{\min}>0$.

Using (\ref{alprime}) and $f'_{\min}$ as well the fact that the coefficient
in front of $\alpha'_{2,3}(v)$ in the expansion is negative and bounded away
from zero, then the $\alpha'_{2,3}(v)$ term can be made as negative as desired
by choosing $\lambda$ sufficiently large.
Meanwhile, the potentially positive remaining term (last two lines) is bounded
from above.
Thus by taking $\lambda$ sufficiently large, it can be ensured that the
$\alpha'_{2,3}(v)$ term is more negative than the remaining term is, perhaps,
positive.
As a result, for such $\lambda$ the expansion is negative.

Taking $\lambda$ sufficiently large is, by definition, taking $\eps_{2,3}$
sufficiently small.
Thus, the condition on $\lambda$ translates to the final upper bound
imposed in the choice of $\eps_{2,3}$.
Hence, $\eps_{2,3}$ is taken to be smaller than all three upper bounds.

It is important to verify that given this $\alpha_{2,3}(v)$, the integral curve
along $w^a$ that is given by (\ref{transition23}) actually reaches\
$v=v_{2,3}$, i.e.\ verify that the transition ends and reaches the region in
spacetime where $w^a=w_3^{\ a}$.
This is the case for precisely the same reason the previous transition ended.
Since during the transition, the component $w^v\equiv w^a\,dv_a$ is positive
and bounded away from zero, and since the other components are bounded and
the integral curve does not tend to $r=0$, then the integral curve reaches
$v=v_{2,3}$.

Setting $\alpha_{2,3}=1$ in (\ref{exp23}), the expansion once $w^a=w_3^{\ a}$
is found to be
\bean
\label{exp3}
\Theta(w_3) = -\frac{R(v)+2\sqrt{\eps R(v)}\cot\theta-2r}{r^2}
-\frac{R'(v)}{R(v)}
\eean
The expansion is negative if $\tan\theta < \frac{\sqrt{\eps R(v)}}{r}$, but
this is satisfied by (\ref{tantheta}).

When $w^a=w_3^{\ a}$, at the end of the transition from $w_2^{\ a}$, 
$r\leq r_{eh}(v) -3\eps$ as a result of the bound on the change in $\tilde r$.
Using $w_3^{\ r}$, it is easy to verify that $\frac{d\tilde r}{dv}$ is negative
if $r_{eh}(v) \geq r + 3\eps$.
In other words, once $w^a=w_3^{\ a}$ it will remain at
$r \leq r_{eh}(v) - 3\eps$ indefinitely.

\subsection{Transition from $w_3^{\ a}$ to $w_4^{\ a}$ : When $r$ is
sufficiently small} \label{details4}

The transition to $w_4^{\ a}$ starts at $v_3$, which is now chosen.
Let $v_3$ be the smallest $v$ such that $R(v_3)\geq \frac{1}{2}R_0$ and
such that $r_3\equiv r(v_3)$ satisfies $r_3\leq 2\eps$.

Along the integral curve of $w_3^{\ a}$, $v$ increases and can reach any
arbitrary large value without the curve getting to the singularity, $r=0$.
This follows since when $r\leq\eps$, $w_3^{\ r}$ is positive, and therefore
$r>\eps$ along the integral curve.

As a result, the curve reaches $v$ such that $R(v)\geq\frac{1}{2}R_0$.
It remains to show that the integral curve along $w_3^{\ a}$ reaches
$r\leq 2\eps$.

It was shown that $r \leq r_{eh}(v) - 3\eps$ holds along the integral curve of
$w_3^{\ a}$.
Since this integral curve reaches arbitrarily large $v$ then it reaches
$v$ large enough so that $R(v)=R_0-\eps$.
It follows that for this $v$ and later, $r\leq R(v)-2\eps$.
The idea is that since $R(v)$ is bounded from above then late enough the
spacetime asymptotes to Schwarzschild and the surface $r=R(v)$ tends to the
event horizon.
Since the integral curve lies at $r\leq r_{eh}(v)-3\eps$ then once the surface
$r=R(v)$ is close enough to the event horizon, the integral curve must
cross it and it now lies inside the surface $r=R(v)$.

It can easily be shown that once $r\leq R(v)-2\eps$, then
$w_3^{\ r}$ is negative, i.e.\ the integral curve starts tending towards
smaller $r$.
For $r$ in the range $2\eps\leq r\leq R(v)-2\eps$, $w_3^{\ r}$ is
negative and bounded away from zero.
As a result, since, in addition, $w_3^{\ v}=2$, then at finite coordinate $v$,
the integral curve reaches $r=2\eps$.
As already discussed, it does not reach the singularity since 
it tends to $r \gtrsim \eps$.

Thus, $v_3$ was chosen above and the choice of $v_{3,4}$ will be discussed
after the expansion is explored.
The transition is taken to be of the usual form and $w_4^{\ a}$ was given
by (\ref{w4}).
This transition is motivated by a desire to keep the final transition,
as simple as possible. 
After finding $l^a$ for the transition, the expansion is evaluated to be

\begin{align}
\label{exp34}
&\Theta\Big(\big(1-\alpha_{3,4}(v)\big) w_3+\alpha_{3,4}(v) w_4\Big)
& \nonumber \\
& \quad\  = \frac{r^2+(\eps-r) R(v)}{Y} \,\alpha_{3,4}'(v)
& \nonumber \\
& \quad\  -\ \frac{\eps+\alpha_{3,4}(v) \big(r -\eps\big)}{Y} R'(v)
& \nonumber \\
& \quad\  -\ \frac{2}{r^2}\sqrt{Y} \cot\theta
& \nonumber\\
& \quad\ 
+ \frac{\big(R(v)-2r\big)\Big(\alpha_{3,4}(v)\big(1+\alpha_{3,4}(v)\big)
\big(r^2+(\eps-r)R(v)\big)-2\eps R(v)\Big)}{2r^2\,Y} &
\end{align}
where $Y=\eps R(v)-\alpha_{3,4}(v)\big(r^2+(\eps-r)R(v)\big)$.
It follows, since $\eps\leq r \leq2\eps$, that $Y>0$ and exploring the four
terms, it is found that all are manifestly either non-positive or negative,
except for the term that goes like $\alpha_{3,4}'(v)$, which, without further
analysis, may be positive.
However, since the $\cot\theta$ term is negative and bounded away from zero
for all $\alpha_{3,4}(v)\in[0,1]$, then as long as
$\alpha_{3,4}'(v)$ is small enough, the expansion remains negative.
As seen before, the derivative of the transition function is
\bean
\alpha_{3,4}'(v)=\lambda\,f'(x)|_{x=\lambda(v-v_3)}
\eean
with $\lambda=(v_{3,4}-v_3)^{-1}$, where $v=v_{3,4}$ is where the transition to
$w_4^{\ a}$ is complete.
Since $f'(x)$ is bounded then $v_{3,4}$ can be chosen sufficiently large so
that $\lambda$ is sufficiently small and the potentially positive
$\alpha_{3,4}'(v)$ term is dominated by the negative $\cot\theta$ term.
Thus, $v_{3,4}$ is now set and as a result, this transition may terminate at
some very large value of $v$.

During the transition, $w^a \approx 2\Big(\frac{\partial}{\partial v}\Big)^a$
and therefore the transition ends, i.e.\ the integral curve
reaches $v=v_{3,4}$.

Once the transition ends, the expansion along the integral curve of 
$w_4^{\ a}$ is obtained by setting $\alpha_{3,4}(v)=1$
and $\alpha'_{3,4}(v)=0$ in (\ref{exp34}).
Following the previous discussion, the expansion is negative in this case
as well.

\subsection{Transition from $w_4^{\ a}$ to $w_5^{\ a}$ : Closing off
as a 2-sphere} \label{details5}

The required conditions for the transition to $w_5^{\ a}$
to start are already satisfied as soon as the transition to $w_4^{\ a}$
ends,\footnote{Indeed, as mentioned earlier, the reason for
$w_4^{\ a}$ in the first place was simply to allow for the final transition to
be as simple as possible.} and therefore the final transition starts at any
$v_4>v_{3,4}$.

The final transition requires a different way of performing the transition.
A choice of $v_{4,5}$ and a transition of the form
$\Big(1-\alpha_{4,5}(v)\Big)w_4^{\ a}+\alpha_{4,5}(v) w_5^{\ a}$ will not work.
The integral curve, in this case, will never reach $v_{4,5}$ since as
$\alpha_{4,5}$ tends to $1$, then $w^v$ tends to 0, but $\alpha_{4,5}(v)$
increases only if $v$ increases and therefore a transition where the
transition function depends on $v$ alone in this way, is not possible.
A simple solution is to take, in this case, a transition function that depends
on $v$ and $\theta$, such as $\alpha_{4,5}(v+\theta)$.

The vector fields $w_4^{\ a}$ and $w_5^{\ a}$ defined in (\ref{w4}) and
(\ref{w5}), respectively, satisfy $w_4^{\ a}w_{4 a}=w_5^{\ a}w_{5 a}$ and
$w_4^{\ a}w_{5a}=0$.
Thus, using $\cos(\alpha_{4,5})$ and $\sin(\alpha_{4,5})$, it is possible to
keep $w^a w_a$ independent of $\alpha_{4,5}$ and this simplifies the
expression for the expansion.
For convenience, then, the transition is taken to be of the form
\bean
\label{transition45}
w^a &=& \cos\Big(\alpha_{4,5}(v+\theta)\Big) w_4^{\ a}
+\sin\Big(\alpha_{4,5}(v+\theta)\Big) w_5^{\ a}
\eean
where in this case, the transition function, $\alpha_{4,5}$ is given by
\bean
\label{alpha45}
\alpha_{4,5}(v+\theta)=\frac{\pi}{2} f\Big(\lambda
\big(v+\theta\ -\ (v_4+\theta_4)\big)\Big)
\eean
and in this transition $\lambda$ is simply chosen to be
\bean
\label{lambda45}
\lambda=\frac{4}{\pi}
\eean
Note that this choice of $\lambda$ is really a specification of
$v_{4,5}+\theta_{4,5}$, where the transition ends.
In this transition $\alpha_{4,5}$ varies from $0$ to $\frac{\pi}{2}$ as
$f$ varies from $0$ to $1$.

The transition in this form does end, i.e.\ the integral curve reaches the
region where $w^a=w_5^{\ a}$. 
First, $v$ and $\theta$ are non-decreasing during the entire transition
and it will be shown that $\theta<\frac{\pi}{2}$ throughout the transition.
Initially, when $\alpha_{4,5}\leq\frac{\pi}{4}$, $w^v$ is positive and
bounded away from zero.
In this case the increase in $v$ guarantees that $\alpha_{4,5}$ will
reach $\frac{\pi}{4}$.
When $\alpha_{4,5}\geq\frac{\pi}{4}$,
$w^\theta$ is positive and bounded away from zero.
Here the increase in $\theta$ guarantees that $\alpha_{4,5}$ will reach
$\frac{\pi}{2}$.
Thus, the combination of $v$ and $\theta$ in this way as well as the
particular choice of $\lambda$, ensure that the integral curve
crosses the transition region.

The null field, $l^a$, for the transition is obtained, and the expansion
is evaluated to be
\bean
\label{exp45}
&& \Theta\Bigg(
\frac{\cos\alpha_{4,5}}{\sqrt{\frac{R(v)}{r}-1}}
\Big(\frac{\partial}{\partial v}\Big)^a
+\frac{\sin\alpha_{4,5}}{r}
\Big(\frac{\partial}{\partial \theta}\Big)^a
\Bigg) \nonumber \\
&& \quad\  = \frac{2(\cos\alpha_{4,5}+
\frac{1}{r}\sqrt{\frac{R(v)}{r}-1} \,\sin\alpha_{4,5})}
{1+\sin\alpha_{4,5}} \,\alpha'_{4,5}
\nonumber \\
&& \quad\  -\ \frac{1-\sin\alpha_{4,5}}{R(v)-r}\ R'(v)
\nonumber \\
&& \quad\  -\ \frac{2}{r}\sqrt{\frac{R(v)}{r}-1}\ 
\frac{1-\sin\alpha_{4,5}}{\cos\alpha_{4,5}} \,\cot\theta
\nonumber \\
&& \quad\  -\frac{2\big(\cos(2\alpha_{4,5})-3\big)r +
\big(5-3\cos(2\alpha_{4,5})\big)R(v)}
{2r^2\Big(\cos\frac{\alpha_{4,5}}{2}+\sin\frac{\alpha_{4,5}}{2}\Big)^2}
\eean

Consider the different terms in the expansion.
The term proportional to $R'(v)$ is always non-positive.
This leaves two terms. A term proportional to $\cot\theta$ and the combination
of the $\alpha'_{4,5}$ term and the remaining term, the last line.

The term proportional to $\cot\theta$ is non-positive provided that
$\cot\theta\geq0$, i.e.\ $\theta\leq\frac{\pi}{2}$.
Since $\theta_4$, the angle at which the final transition starts, satisfies
$\theta_4=\theta_{1,2}\leq\frac{\pi}{8}$ then if the additional change in
$\theta$ during the final transition is not greater than, say,
$\frac{\pi}{4}$, then this term will remain non-positive throughout the
transition.
A simple bound on the change in $\theta$ is obtained as follows
\bean
\Delta\theta\equiv \theta_{4,5}-\theta_4 <
\frac{1}{\lambda}\,\lambda\Big(v_{4,5}+\theta_{4,5}-(v_4+\theta_4)\Big)
= \frac{1}{\lambda} = \frac{\pi}{4}
\eean
since at the end of the transition, the argument of $f$ in
(\ref{alpha45}) is equal to $1$, and $v_{4,5}>v_4$.
Thus, $\lambda$ was chosen to ensure that the $\cot\theta$ term remains
non-positive throughout the final transition.

Consider next the combined two terms consisting of the term proportional to
$\alpha'_{4,5}$ and the remaining term, the last line of (\ref{exp45}).
First, since $f'(x)$ is bounded from above, then let $f'_{\max}$ be its
maximum.
From the definition of $\alpha_{4,5}$, (\ref{alpha45}), if follows that
\bean
\alpha_{4,5}'\equiv \alpha_{4,5}'(v+\theta)
= \frac{\pi}{2}\,
f'\Big(\lambda\big(v+\theta\ -\ (v_4+\theta_4)\big)\Big)\ \lambda
\leq 2f'_{\max}
\eean
The combined two terms can be bounded from above as follows
\bean
\label{combo45}
&& \frac{2(\cos\alpha_{4,5}+
\frac{1}{r}\sqrt{\frac{R(v)}{r}-1} \,\sin\alpha_{4,5})}
{1+\sin\alpha_{4,5}} \,\alpha'_{4,5}
\nonumber \\
&& -\ \frac{2\big(\cos(2\alpha_{4,5})-3\big)r +
\big(5-3\cos(2\alpha_{4,5})\big)R(v)}
{2r^2\Big(\cos\frac{\alpha_{4,5}}{2}+\sin\frac{\alpha_{4,5}}{2}\Big)^2}
\nonumber \\
< && 4\Bigg(1+\frac{1}{r}\sqrt{\frac{R(v)}{r}-1}\,\Bigg)f'_{\max}
-\frac{R(v)}{4r^2} \nonumber \\
< && 4\Bigg(1+\frac{1}{r}\sqrt{\frac{R_0}{r}-1}\,\Bigg)f'_{\max}
-\frac{R_0}{8r^2}
\eean
where, in the first inequality, $R(v)-4r\geq\frac{1}{2}R_0-8\eps\geq 0$, which 
follows from (\ref{parmeps}), was used.
The last inequality follows since $\frac{1}{2}R_0\leq R(v)\leq R_0$.

Given $R_0$ and $f'_{\max}$ it follows that there exists some positive value
$\eps^*$ such that if $r\leq 2\eps^*$ then the right hand side of
(\ref{combo45}) is negative.
Therefore, $\eps^*$, the parameter appearing in (\ref{parmeps}), is
now chosen\footnote{Notice how the choice of $\eps^*$, for
example, depends only on $R_0$ and $f'_{\max}$. Even though this choice is 
discussed in this section, after the choices of 
$\theta^*_1,\theta^*_2 \text{, and } \theta^*_3$ have been made, it clearly
does not depend on any of these choices.}
and since by (\ref{parmeps}), $\eps\leq\eps^*$, then this ensures that the
contribution to the expansion from the combined two terms is negative.
Since the other terms in the expansion are non-positive, it follows that the
expansion throughout the final transition is always negative. 

Once $w^a=w_5^{\ a}$ the integral curve continues along
$(\frac{\partial}{\partial \theta})^a$
until the axis is reached at $\theta=\pi$.
Along the integral curve of $w_5^{\ a}$, the expansion is that of a
2-sphere, as given by (\ref{exp2sphere}).
It is negative, since, in this region, $r<R(v)$.
The ending point, $\theta=\pi$, of the integral curve, is the south pole of
the 2-surface. 
The 2-surface obtained by translation of the curve by the axial Killing field
is smooth there, since, in a neighborhood of this point, the 2-surface is
just a portion of a 2-sphere.

\subsection{The 2-surface obtained in this way}

Since the integral curve starts and ends at the axis, then the resulting
2-surface is compact.
Since $w^a$, and the axial Killing field are spacelike everywhere, then
so is the resulting 2-surface.
The 2-surface is smooth since $w^a$ is smooth everywhere and at the north
pole, $\theta=0$, and at the south pole, $\theta=\pi$, the surface was shown
to be smooth.
Finally, the expansion of outgoing null geodesics orthogonal to it
is negative everywhere, as was shown throughout the construction.
It is, therefore, an outer trapped surface and it contains the initial point.
This completes the proof.


\section{Trapped surfaces} \label{trappedsurfaces}

The situation regarding outer trapped surfaces in Vaidya spacetimes is now
clear.
These can lie arbitrarily close to the event horizon in any region of the
spacetime.
What about trapped surfaces, i.e.\ surfaces where both expansions are required
to be negative?
Exploring the region containing outer trapped surfaces and its boundary leads
to a similar question about the region containing trapped surfaces.

Future trapping horizons \cite{Hayward94} and dynamical horizons
\cite{AshtekarKrishnan} are foliated by inner trapped and marginally outer
trapped surfaces, i.e.\ foliated by surfaces that are almost trapped, except
that in the outgoing direction the expansion vanishes instead of being
everywhere negative.
It is the region containing trapped surfaces and not the one containing
outer trapped surfaces that is relevant when trying to explore where
future trapping horizons and dynamical horizons can exist.

It will now be shown that in Vaidya spacetimes that contain a flat region, 
a portion of the event horizon does not have any trapped surfaces lying close
to it.\footnote{This result, as well as the argument that proves it,
were suggested by Bob Wald.}
As a result, the boundary of the region containing trapped surfaces is not,
in general, the event horizon.

Consider some spacelike 2-surface, $S$, in spacetime.
Let this 2-surface be embedded in a spacelike 3-surface, $\Sigma$, which,
itself, is embedded in the four dimensional spacetime.
Let $s^a$ be the outgoing spacelike unit normal to $S$ in $\Sigma$, and let
$t^a$ be the timelike, future directed, unit normal to $\Sigma$ in
the spacetime.
The outgoing null normal to the 2-surface, is then given by
\bean
l^a=t^a+s^a
\eean
The expansion in this case is given by
\bean
\label{expparts}
\Theta=q^{ab}\nabla_a l_b =h^{ab}\nabla_a t_b - s^a s^b\nabla_a t_b
+q^{ab} D_a s_b
\eean
where $q^{ab}$ is the inverse of the induced metric on $S$, $h^{ab}$ is the
induced metric on $\Sigma$, and $D_a$ is the covariant derivative
associated with $\Sigma$.
If instead of the outgoing direction, the ingoing one is of interest, then
this derivation can be repeated with the change $s^a\to -s^a$.
The first two terms in (\ref{expparts}) include the trace of the extrinsic
curvature of the 3-surface, $\Sigma$, in spacetime and the component of
the extrinsic curvature orthogonal to $S$.
These are independent of the change $s^a\to -s^a$.
The last term in (\ref{expparts}), the trace of the extrinsic curvature of
the 2-surface, $S$, in the 3-surface, $\Sigma$, changes sign with the
change $s^a\to -s^a$.

This can now be applied in flat space in inertial coordinates.
Let $\Sigma$ be a $t=\text{const}$ hyperplane.
Then $\Sigma$ is a 3-surface with vanishing extrinsic curvature in the flat
four dimensional spacetime.
As a result, any 2-surface in this hyperplane will have the two expansions 
of equal size and of opposite
signs (or, of course, both vanishing).\footnote{This follows since, in this
case the first two terms in (\ref{expparts}) vanish and the last term changes
sign between the ingoing and outgoing expansions.}
If some 2-surface embedded in such a hyperplane in flat space is
inner trapped, with that expansion being everywhere negative, then this
2-surface is not outer trapped, with that expansion being everywhere positive,
and vice versa.

For a trapped surface to extend into the flat region of a Vaidya spacetime,
it will be shown that it needs to ``bend down in time'', since otherwise,
the expansions cannot both be negative.
An example of a surface that ``bends down in time'' is the intersection of
the past lightcones of two single events in flat space.
This spacelike 2-surface has both ingoing and outgoing expansions
everywhere negative.
It is not, however, a trapped surface, because, as the intersection of
two past lightcones, this surface is not compact.

\begin{figure}
      \begin{center}
    \resizebox{13cm}{!}{\includegraphics{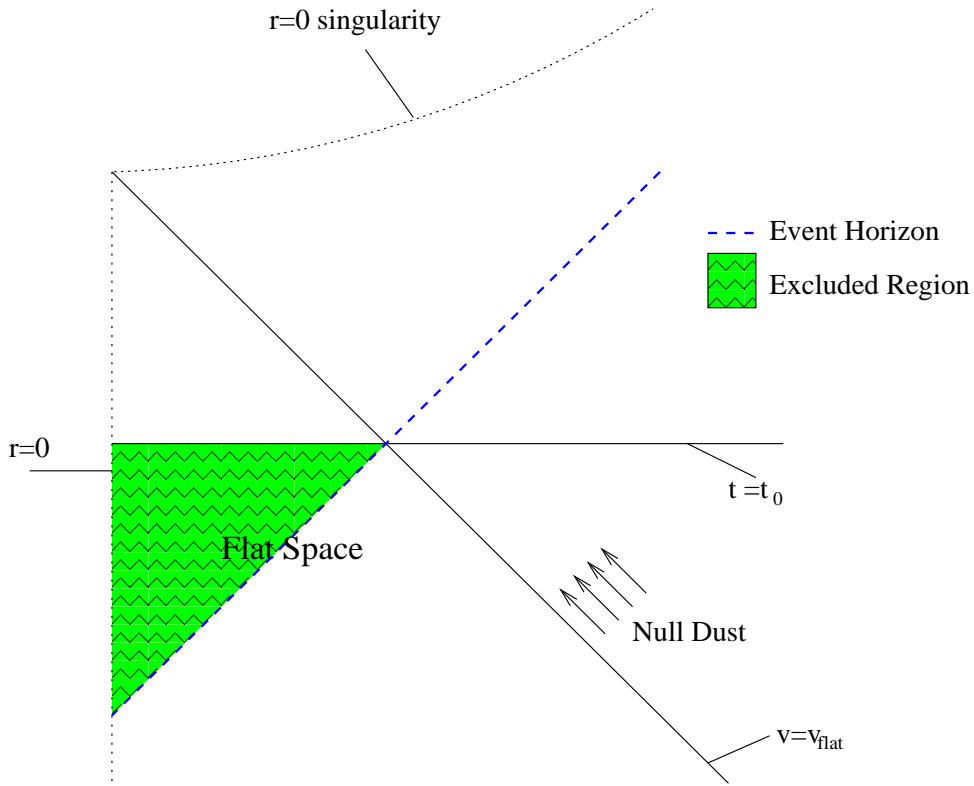}}
  \end{center}
  \caption{A spacetime diagram of the collapse of null dust in flat space.
The coordinate $t$ is such that $(\frac{\partial}{\partial t})^a$ is
pointing upwards in the flat region in the diagram.
The surfaces $v=v_\text{flat}$ and $t=t_0$ are shown. The excluded region is
the region inside the event horizon satisfying $t<t_0$. This is the shaded
triangular region in the diagram. It follows from the proposition that trapped
surfaces cannot extend into this region. As a result, the boundary of the
region containing trapped surfaces, is not, in general, the event horizon.}
  \label{ExcludedFig}
\end{figure}

It will be shown that because of the required ``bending down in time'',
a trapped surface, which extends into the flat region of the spacetime,
cannot have a minimum there, for any inertial time $t$.
It will follow that there is an excluded region, where not having a minimum
for any inertial time $t$ cannot occur without the surface crossing the event
horizon, which is impossible.
It will therefore be shown, that no trapped surfaces can pass in this
excluded region.
The situation is shown in Fig.\ \ref{ExcludedFig}.
The excluded region can be seen in the flat region of the spacetime,
bordering the event horizon.
The proposition is now stated and proved.

{\bf Proposition}: Consider a Vaidya spacetime with metric as
in (\ref{Metric}) such that $R(v)$ is a non-negative, non-decreasing,
bounded function, and such that $R(v)=0$ for $v\leq v_\text{flat}$ and 
$R(v)>0$ for $v>v_\text{flat}$.
Let $t$ be the global inertial time coordinate in the flat region such that
the 2-spheres of the global spherical symmetry of the Vaidya spacetime are
at rest.
Let $t_0$ be the value of $t$ at the intersection of the event horizon with
the surface $v=v_0$.
Then, no trapped surface in this spacetime may contain a point with $v<v_0$
and $t<t_0$.

{\bf Proof}: Let $U$ be the region in this spacetime that contains all the
points that lie inside, or coincide with, the event horizon and
with $t\leq t_0$. Thus, in the figure, this is the shaded region including
the portion of the 3-surface $t=t_0$ at the top of the region, and including
the portion of the event horizon at the bottom of this region.
The region $U$ is compact.

Assume that a trapped surface contains some point $(t_1,x_1,y_1,z_1)$ that
lies strictly inside this region, i.e.\ $t_1<t_0$, and, of course, it is
strictly inside the event horizon.
Let $W$ be the intersection of the trapped surface with $U$.
The region $W$ is compact and it is non-empty, since, by assumption,
it contains the point above.
As a result a minimum for $t$ exists in $W$.
Let $(\tilde t,\tilde x,\tilde y, \tilde z)$ be
a point in $W$ where this minimum for $t$ is attained.

This point cannot lie in the boundary of $U$, since this would mean that
either $\tilde t=t_0$ or that the point lies in the event horizon.
The first is impossible, since, by assumption, another point in $W$
satisfies $t_1<t_0$, and therefore this would contradict this point being a
minimum for $t$ in $W$.
The second is impossible, since trapped surfaces lie strictly inside the
event horizon, so the point cannot lie on the event horizon.
Therefore, the minimum for $t$ is attained strictly inside $U$, and this
implies that $(\tilde t,\tilde x,\tilde y, \tilde z)$ must be a local minimum
for $t$.

Without loss of generality, the Cartesian coordinates can be chosen so that
at $(\tilde t,\tilde x,\tilde y, \tilde z)$, the $z$-axis is orthogonal to
the trapped surface.
It follows that in a neighborhood of this point, the 2-surface can be expressed
in terms of $x$ and $y$.
Therefore, in a neighborhood of this point the surface's coordinates are
given by
\bean
\label{trapped2surface}
\Big(T(x,y),x,y,Z(x,y) \Big)
\eean
with $T(x,y)$ and $Z(x,y)$ some smooth functions in this neighborhood.
Since $(\tilde t,\tilde x,\tilde y, \tilde z)$ is a local minimum for $t$ then
\bean
\label{tprime}
\frac{\partial T(x,y)}{\partial x}|_{(\tilde x,\tilde y)}=0
\quad\quad \text{and} \quad\quad
\frac{\partial T(x,y)}{\partial y}|_{(\tilde x,\tilde y)}=0
\eean
Similarly, since the $z$-axis is orthogonal to the 2-surface at this point
then
\bean
\label{zprime}
\frac{\partial Z(x,y)}{\partial x}|_{(\tilde x,\tilde y)}=0
\quad\quad \text{and} \quad\quad
\frac{\partial Z(x,y)}{\partial y}|_{(\tilde x,\tilde y)}=0
\eean

Consider the 3-surface given by $t-T(x,y)=0$.
At $(\tilde t,\tilde x,\tilde y, \tilde z)$, the normal to the surface,
$\nabla_a\Big(t-T(x,y)\Big)$, is timelike.
Since this gradient is smooth in the coordinates, then it is timelike in some
small neighborhood of $(\tilde t,\tilde x,\tilde y, \tilde z)$.
Furthermore, in this neighborhood the 2-surface given by
(\ref{trapped2surface}) is embedded in this 3-surface.
Consider the sum of the outgoing and ingoing expansions at the point
$(\tilde t,\tilde x,\tilde y, \tilde z)$.
Since, in the interchange of ingoing and outgoing, $s^a$ changes sign in
(\ref{expparts}), then in this sum only the first two terms of
(\ref{expparts}) will contribute.
These two terms depend on $t^a$, the future directed timelike unit normal to
the 3-surface, in a neighborhood of the point.
They depend on $s^a$, the spacelike unit normal to the 2-surface in the
3-surface, as well as on $h^{ab}$, the metric on the 3-surface.
However, in order to evaluate the sum of the expansions at
$(\tilde t,\tilde x,\tilde y, \tilde z)$,
$s^a$ and $h^{ab}$ are needed only at that point.

A simple evaluation shows that
\bean
t^a &=& \frac{1}{\sqrt{1-T_x^{\ 2}-T_y^{\ 2}}}
\ (\frac{\partial}{\partial t})^a \nonumber \\
&+& \frac{T_x}{\sqrt{1-T_x^{\ 2}-T_y^{\ 2}}}
\  (\frac{\partial}{\partial x})^a
+ \frac{T_y}{\sqrt{1-T_x^{\ 2}-T_y^{\ 2}}}
\ (\frac{\partial}{\partial y})^a
\eean
where $T_x\equiv\frac{\partial T(x,y)}{\partial x}$ and similarly for $T_y$.
At $(\tilde t,\tilde x,\tilde y, \tilde z)$, using (\ref{tprime}),
the metric of the 3-surface is given by
\bean
h_{ab}|_{\big(\tilde x,\tilde y,Z(\tilde x,\tilde y)\big)} =
dx_a dx_b + dy_a dy_b + dz_a dz_b
\eean
Finally, using (\ref{tprime}) and (\ref{zprime}), the normal to the 2-surface
in the 3-surface at $(\tilde t,\tilde x,\tilde y, \tilde z)$ is given
by\footnote{This $s^a$ may
be the outgoing or the ingoing normal to the 2-surface and without the entire
2-surface, this remains undetermined. However, since it is the sum of the two
expansions that is of interest and in it only $s^a s^b$ appears, then, in this
case, the sign of $s^a$ does not matter.}
\bean
s^a|_{(\tilde x,\tilde y)} = (\frac{\partial}{\partial z})^a
\eean
It is now possible to use (\ref{expparts}) to evaluate the sum of the two
expansions.
Using $t^a$, $h_{ab}$, and $s^a$ as above, as well as (\ref{tprime}),
the sum of the two expansions is found to be
\bean
\label{condpartial}
\big(\Theta_1 +\Theta_2\big)|_{(\tilde t,\tilde x,\tilde y, \tilde z)}
= 2(\frac{\partial^2 T(x,y)}{\partial x^2} +
\frac{\partial^2 T(x,y)}{\partial y^2})
\eean
Since the 2-surface is trapped, then the sum of the two expansions is negative,
and therefore at least one of the two terms in (\ref{condpartial})
must be negative.
This is a contradiction to $(\tilde t,\tilde x,\tilde y, \tilde z)$ being
a local minimum. (It also shows that for a surface to be trapped, it has
to ``bend down in time'' everywhere in the flat region.)

Thus, the assumption that $(t_1,x_1,y_1,z_1)$, a point of the trapped surface,
lies strictly inside $U$, cannot hold.
This completes the proof.


\section{Discussion} \label{discussion}

In Vaidya spacetimes, the situation regarding outer trapped surfaces,
is now clear.
Outer trapped surfaces can reach arbitrarily close the the event horizon
everywhere and Eardley's conjecture is true for these spacetimes.

Extending the main result to other spacetimes using a similar procedure,
seems unlikely since the technique used here relies on precise control of the
location of the integral curve relative to the spherically symmetric apparent
3-horizon and the event horizon.
The level of precision, \emph{required for this particular method},
can be obtained in Vaidya spacetimes in the ingoing Eddington-Finkelstein
coordinates, but appears to be substantially harder to obtain in general
spherically symmetric spacetimes and much more so in spacetimes that are
not spherically symmetric.
Nonetheless, the class of spacetimes covered by the main result is wide as it
includes spacetimes that start flat and later, with some collapsing matter,
form a black hole.
As a result, since it was shown that outer trapped surfaces exist even in the
flat region in such spacetimes, then this appears to, perhaps, capture the
essential features of general black hole collapse spacetimes.
This then strengthens the expectation that Eardley's conjecture is true, in
general.

The situation regarding trapped surfaces in Vaidya spacetimes was explored
as well.
A proposition was proved, showing that there is a portion of the flat region
of a Vaidya spacetime that is excluded, i.e.\ that trapped surfaces cannot
enter this region.
This is consistent with the results of Schnetter and
Krishnan \cite{SchnetterKrishnan}.
They describe finding a marginally trapped surface that extends into the flat
region of a Vaidya spacetime.
Since the flat region is not excluded entirely, it follows that the surface
described by Schnetter and Krishnan can extend into the flat region in the
non-excluded part.

It would be of interest to find in Vaidya spacetimes the exact location of the
boundary of the region containing trapped surfaces, as this will give a better
understanding of where can surfaces such as future trapping horizons and
dynamical horizons be located in these spacetimes.
In Fig.\ \ref{ExcludedFig}, this boundary must lie somewhere above the $t=t_0$
surface.

\bigskip


\noindent
{\bf Acknowledgments}

\medskip

I am greatly indebted to my advisor, Bob Wald, for many useful discussions
and guidance along the way.
I have also benefited from discussions with Akihiro Ishibashi.
This research was supported in part by NSF grant PHY-0456619 to
the University of Chicago.
Submitted in partial fulfillment of the requirements of the degree of 
Doctor of Philosophy, University of Chicago, Chicago, Illinois.


\end{document}